\documentclass[intlimits,twoside,a4paper]{article}

\usepackage[cp1251]{inputenc}
\usepackage[normalem]{ulem} 
\usepackage[eqsecnum]{cmpj3}

\usepackage{bm}

\issue{2023}{26}{3}{33901}
\doinumber{10.5488/CMP.26.33901}

\title[How should a small country respond to climate change?]
{How should a small country respond to climate change?}

\author{A. D. J. Haymet\orcid{0000-0002-5617-2106}\refaddr{label1,label2}
}
\addresses{
	\addr{label1} Antarctic Science Foundation 203 Channel Highway Kingston, Tasmania, Australia
\addr{label2} Scripps Institution of Oceanography, University of California San Diego, San Diego, CA, USA
}

\Keywords{climate change, greenhouse gases, small country response}

\date{Received April 12, 2023, in final form April 23, 2023}
\begin{document}

\maketitle

\begin{abstract}
Responses to the global climate crisis often focus on the largest current emitters of greenhouse
gases. However, analysis shows that about a third of emissions come from a collection of small
emitters, each contributing one- to two-percent of the total additional CO$_2$ injected into the
communal atmosphere. Attempts to hold global warming to less than 1.5\textcelsius~ cannot succeed without also
reducing emissions from these small countries.
\printkeywords
\end{abstract}


\maketitle



What should small countries deeply affected by GHG (greenhouse gas) emissions do? How do they convince large countries to drastically reduce emissions, especially if they themselves mine the products which --- when used --- greatly add to global warming?  For example, to save the Great Barrier Reef, a major tourist attraction and biodiverse ecosystem, Australia must convince the northern hemisphere to stop emitting CO$_2$ and other heat-trapping greenhouse gases as quickly as possible. Since 1955, by actual measurements up to December 2022 (not ``calculations'' or ``theory'') humans have dumped over 345 ($\pm 2$) zettajoules of extra heat into the earth system \cite{1}, almost all of it into the ocean. That extra heat fuels the weather every single day. It perverts our winds and rainfall. It supercharges cyclones. It bleaches coral reefs. It causes sea-level rise. It disturbs our fisheries and agriculture. And it creates a higher probability of extreme bushfire events.

There is no doubt that the damage comes from this excess heat.   So where does the heat come from? That is also no mystery. Fortunately, since 1965 (at least) we have known exactly. We humans have increased the atmospheric CO$_2$ concentration by 50\%, to over 410 ppm. That CO$_2$ plus methane and other greenhouse gases (GHG) (assisted by some black soot) trap more heat than nature intended in the atmosphere, and that heat is rapidly transferred to the ocean, due to its massively higher heat capacity. Since 1958, no other explanation has survived scientific scrutiny. 

About 92.5\% of the heat stays in the ocean \cite{2}. Only about 2\% stays in the atmosphere, but since that is where we live, we tend to focus on ``global warming'' of the atmosphere. The massive heating of the ocean is much more important and much more damaging. Additional small percentages of that heat melt glaciers formerly grounded on land, and also heat up the soil and land mass directly.

In many ways it is a great relief that we know precisely what is causing this extra heat. Because knowledge is power. We can reverse it, albeit agonizingly slowly. Imagine if our scientific knowledge of ocean warming was not so precise! We would be suffering in terrible uncertainty about how to respond to the current crisis.

But we know exactly what to do. To save Australia's fisheries and farms, its reefs and its forests and its very lifestyle, we need to stop the production of this extra heat, and we can do that by steadily quickly reducing the global output of greenhouse gases, and eventually become ``carbon neutral''.

Thanks to scientist Susan Solomon \cite{3}, we have known for a decade how long this reversal takes. It is agonizingly long. That which we sullied in 50--100 years takes a thousand years to undo. But that is NOT an argument to do nothing, but rather an exhortation --- a commandment if you will --- to start reductions today. Many jurisdictions, including Sweden and California, and major companies all aim to reach carbon neutrality by 2030. Australia is blessed with copious renewable resources, more than enough to make hydrogen from renewables, and superb engineering \& scientific talent. We are indeed the ``lucky country'' in responding to this excess heat challenge.

Imagine if we had acted upon the rigorous scientific reports of 1965, the Lyndon Johnson ``President’s Science Advisory Committee Report on Atmospheric Carbon Dioxide'' \cite{4}, and the 1979 report chaired by Charney \cite{5}, and had steadily reduced our GHG emissions starting 55 years ago. Our current observed global warming of 1.1\textcelsius{} would be vastly reduced, and our time to reverse the harm vastly shorter. The currently observed bias towards formerly unlikely disastrous events would be much reduced. Alas we cannot go back and implement their wisdom over the last 55 years. But we can start today. In another 55 years, the next two generations will at least be thanking us for starting today, as much as they will curse us for not acting in 1965.

Why is \uline{the job} of Australia to convince the northern hemisphere to reduce their GHG output? Because their pollution eventually reaches us. The data in figure~\ref{fig2} below, documents the difference in CO$_2$ concentration between northern and southern hemispheres, since 1958. 

\begin{figure}[!t]
	\centerline{\includegraphics[width=0.8\textwidth]{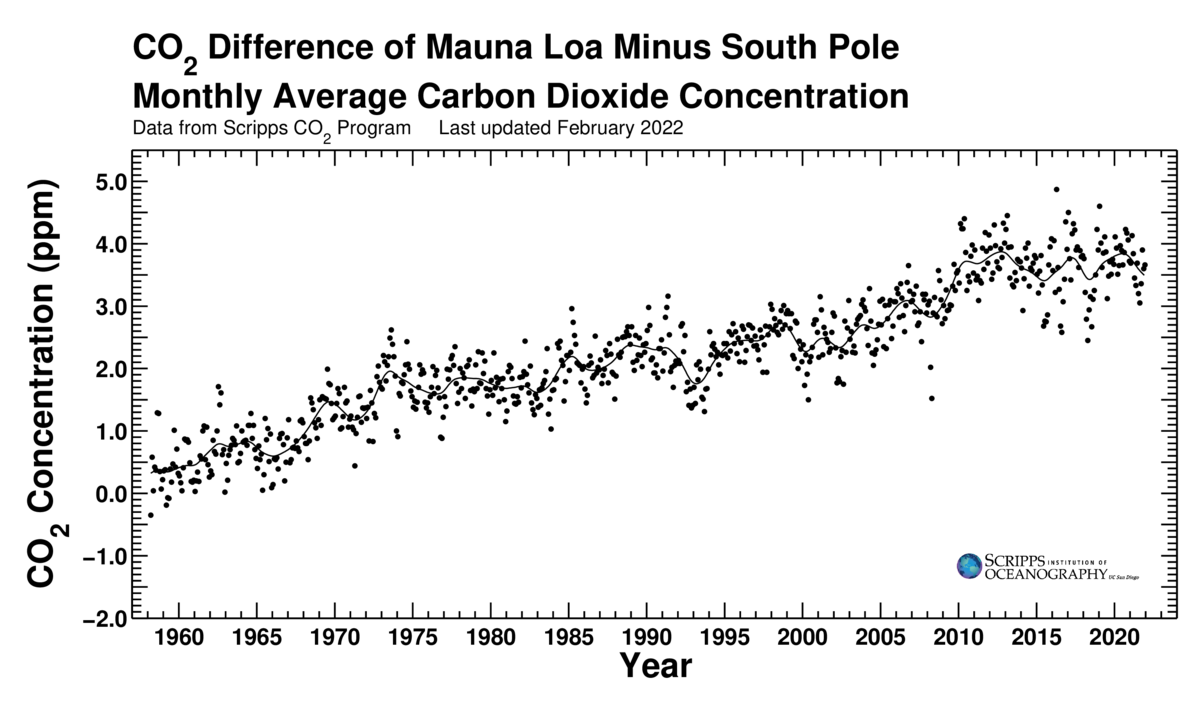}}
	\caption{(Colour online) Pollution from the northern hemisphere. Black dots: difference of the monthly average atmospheric carbon dioxide concentration at Mauna Loa Observatory, Hawaii and South Pole, Antarctica versus time where CO$_2$ concentration is in parts per million (ppm). Black curve: spline fit to the CO$_2$ data. Data from Scripps CO$_2$ Program. URL~\url{https://scrippsco2.ucsd.edu/graphics_gallery/mauna_loa_and_south_pole/mauna_loa_and_south_pole_difference.html} }
	\label{fig1}
\end{figure}
\begin{figure}[!t]
	\centerline{\includegraphics[width=0.7\textwidth]{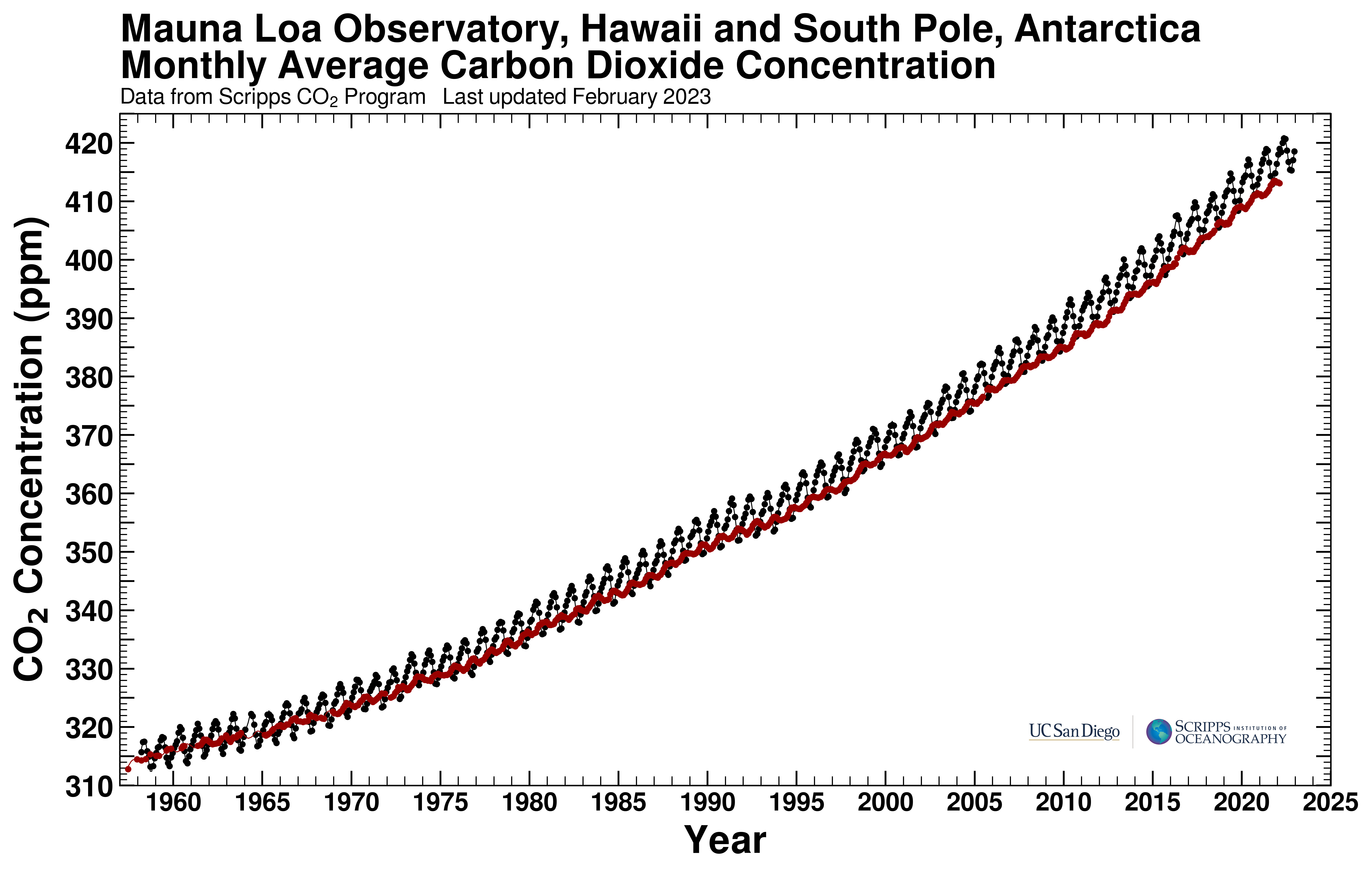}}
	\caption{(Colour online) Black curve: Monthly average atmospheric carbon dioxide concentration versus time at Mauna Loa Observatory, Hawaii (20$^\circ$N, 156$^\circ$W) where CO$_2$ concentration is in parts per million in the mole fraction (ppm). Monthly data are shown as dots and connected with straight lines. The record is distinguished by its pronounced seasonal cycle. Red Curve: Monthly average atmospheric carbon dioxide concentration versus time at the South Pole, Antarctica, where CO$_2$ concentration is in parts per million (ppm). Monthly data points are shown as dots and connected approximately by a smooth curve. Data from Scripps CO$_2$ Program. URL~\url{https://scrippsco2.ucsd.edu/graphics_gallery/mauna_loa_and_south_pole/mauna_loa_and_south_pole.html }}
	\label{fig2}
\end{figure}
\begin{figure}[!t]
	\centerline{\includegraphics[width=0.5\textwidth]{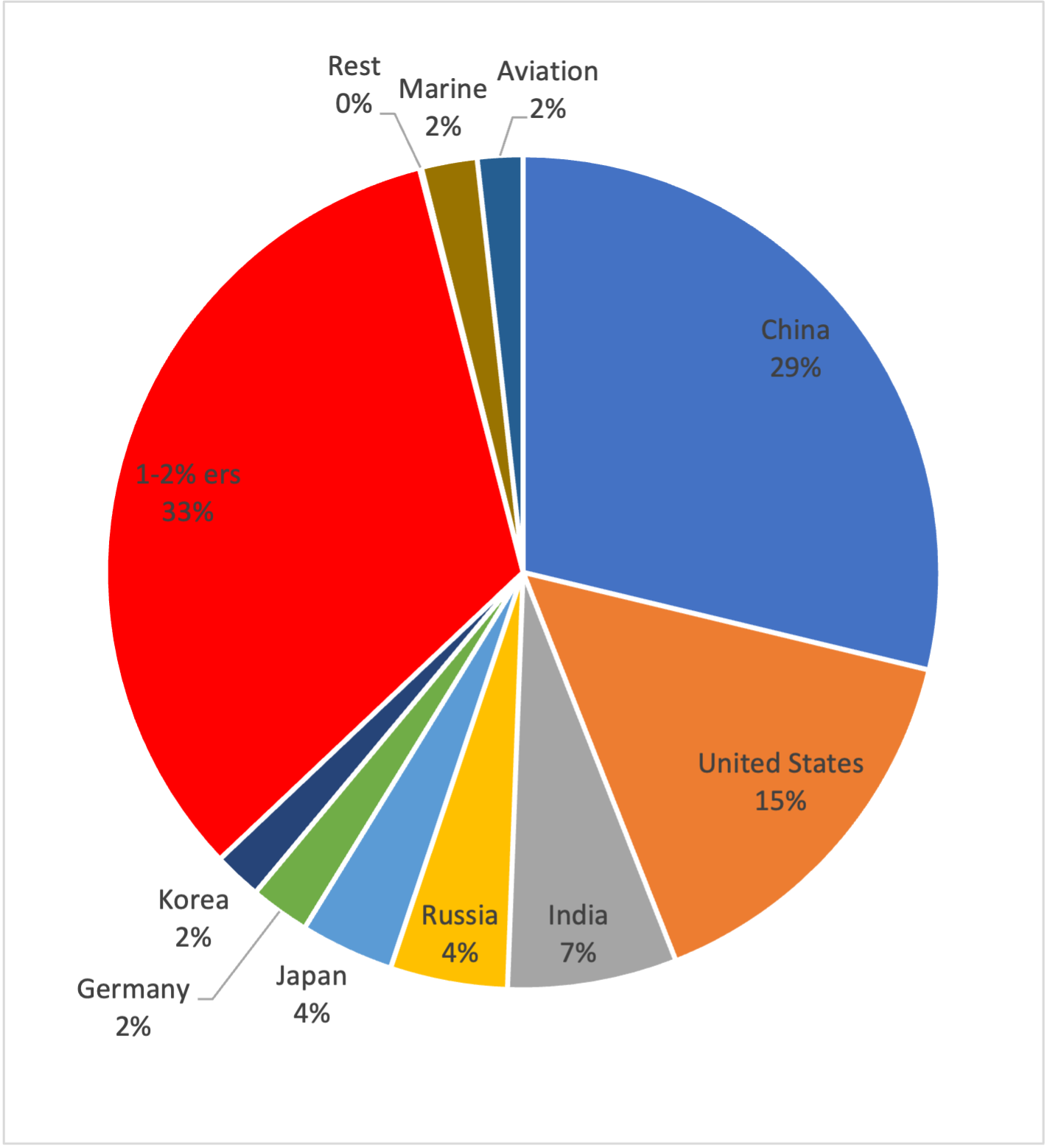}}
	\caption{(Colour online) Annual CO$_2$ emissions from fuel combustion by country or country group. From IEA data. }
	\label{fig3}
\end{figure}
Figure~\ref{fig1} arises from the famous Keeling CO$_2$ measurements at Mauna Loa Observatory, Hawaii, combined with the South Pole observatory in Antarctica, shown in figure~\ref{fig2}.
In figure~\ref{fig2}, both data sets increase over the decades, due to human emissions, but the northern hemisphere is increasing faster. It takes quite a few years now for the extra CO$_2$ in the northern hemisphere to make it across the natural atmospheric barriers at the equator to the southern hemisphere.  How should Australia go about this vital task?

Australia has a choice of two basic tactics to get the northern hemisphere to stop emitting:
\begin{itemize}
\item ``Do as I do'',
\item ``Do as I say''.
\end{itemize}

\textbf{``Do as I do''} means setting aggressive reduction targets for Australia, including the consequential emissions from its exports. It means avoiding accounting tricks from a long-forgotten late-night bargaining session in Kyoto by former Australian Environment Minister Robert Hill’s righthand person.

\textbf{``Do as I say''} means... well, what does it mean? How does one go about convincing the many dozens of global emitting countries to do what Australians themselves are unwilling to do? As shown in figure~\ref{fig2}, far from reducing global emissions, last year GHG emissions increased and a record rate. We are making a horrendous problem even worse, each day, every day, even though we know exactly how to fix it. We can make electricity without generating CO$_2$ or releasing methane. The same for transportation. And farming practices. We still have work to do in metallurgy and cement manufacture and in other industrial processes, but we have all the ingenuity and motivation we need.

Just how many countries does Australia have to convince? A lot. More than one third of total global emissions come from countries who are ``one-to-two percent'' contributors, just like Australia. Roughly 40 countries emit between 2\% and 0.1\% of the global total. That third cannot be ignored, even if strong action were to be taken by the current six heaviest emitters China, USA, India, Russia, Japan and Germany (60\% of total emissions). Consideration of emissions per capita, and accumulated historical emissions per capita, rather than just current total per country, lead to an even greater increased responsibility for Australia.

\section*{Conclusions}
Over the last 15 years there have been some great Foreign Ministers of Australia. Sometimes they have been sent out to do their job with proven mechanisms in place to reduce Australia's CO$_2$ pollution, like the emissions reduction scheme former PM Julia Gillard promised in the 2010 election, and legislated successfully. At other times Australia has sent Foreign Ministers out into the world with their arms tied behind their back. Australia must have a consistent and effective Australian policy, and a good time to start is --- now.
\vspace{1cm}
\newline
\textit{Tony Haymet PhD FTSE is a scientist, business founder and owner, and Chair of the Antarctic Science Foundation. He has served as Chief of CSIRO Marine \& Atmospheric Research, and Director of Scripps Institution of Oceanography.}

\ukrainianpart

\title{Як малій країні слід реагувати на зміни клімату?}
\author{А. Д. Д. Хаймет\refaddr{label1,label2}}

\addresses{
	\addr{label1} Антарктична наукова фундація, 203 Ченнел Хайвей, Кінгстон, Тасманія, Австралія
	\addr{label2} Океанографічний інститут Скриппса, Каліфорнійський університет в Сан-Дієго, Сан Дієго, Каліфорнія, США
}

\makeukrtitle

\begin{abstract}
Говорячи про глобальну кліматичну кризу, часто зосереджуються на країнах, які на теперішній час викидають найбільшу кількість 
парникових газів. Однак аналіз показує, що близько третини викидів зумовлюють малі країни, 
на долю кожної з яких припадає від одного до двох відсотків 
загального надлишкового CO$_2$, що викидається в атмосферу. 
Спроби втримати глобальне потепління в межах 1.5\textcelsius~ не можуть бути успішними без зменшення викидів цих малих країн.
	\keywords зміни клімату, парникові гази, реакція малої країни
\end{abstract}

\lastpage
\end{document}